\documentclass[prb,twocolumn,showpacs]{revtex4}
\usepackage{amsmath}
\usepackage{amssymb}
\usepackage{bm}
\usepackage{epsfig}
\usepackage{graphicx}
\usepackage{color}
\usepackage[english]{babel}

\renewcommand{\Re}{\mathop{\rm Re}}
\renewcommand{\Im}{\mathop{\rm Im}}

\begin{document}

\newcount\timehh  \newcount\timemm
\timehh=\time \divide\timehh by 60
\timemm=\time
\count255=\timehh\multiply\count255 by -60 \advance\timemm by \count255

\title{
Generation and detection of mode-locked spin coherence in (In,Ga)As/GaAs quantum dots by laser pulses of long duration}
\author{S. Spatzek$^{1}$, S. Varwig$^{1}$, M.~M. Glazov$^{2}$, I.~A. Yugova$^{1,3}$, D.~R. Yakovlev$^{1,2}$, D. Reuter$^{4}$, A.~D. Wieck$^{4}$,  and M. Bayer$^{1,2}$}

\affiliation{$^1$ Experimentelle Physik 2, Technische Universit\"at Dortmund, 44221 Dortmund, Germany}

\affiliation{$^2$ A. F. Ioffe Physical-Technical Institute, Russian
Academy of Sciences, 194021 St. Petersburg, Russia}

\affiliation{$^3$ Department of Solid State Physics, Physical
Faculty of St.Petersburg State University, 198504 St. Petersburg,
Russia}

\affiliation{$^4$ Angewandte Festk\"orperphysik, Ruhr-Universit\"at
Bochum, D-44780 Bochum, Germany}

%\date{\today, file = \jobname.tex, printing time = \number\timehh\,:\,\ifnum\timemm<10 0\fi \number\timemm}

\begin{abstract}
Using optical pulses of variable duration up to 80 ps, we report on
spin coherence initialization and its subsequent detection in
$n$-type singly-charged quantum dots, subject to a transverse
magnetic field, by pump-probe techniques. We demonstrate
experimentally and theoretically that the spin coherence generation
and readout efficiencies are determined by the ratio of laser pulse
duration to spin precession period: An increasing magnetic field
suppresses the spin coherence signals for a fixed duration of pump
and/or probe pulses, and this suppression occurs for smaller fields
the longer the pulse duration is. The reason for suppression is the
varying spin orientation due to precession during pulse action.
\end{abstract}


\maketitle

\section{Introduction}

Research on spin coherence generation, manipulation and detection
has become a topical area in semiconductor
physics.\cite{awschalom_book,spin_phys} A prospective system to
study spin coherence is an ensemble of $n$-type quantum dots (QDs).
Short optical pulses can induce efficiently long-living spin
coherence in such structures which subsequently can be traced by
precession about an external magnetic
field.\cite{PhysRevLett.94.227403,greilich06} Due to the combined
action of a periodic train of pump pulses and the hyperfine
electron-nuclear interaction, a mode-locking of electron spin
precession may
occur.\cite{A.Greilich07212006,A.Greilich09282007,carter:167403} As
a result, an ensemble of about a million QD electron spins is pushed
into a regime given by a limited number of precession modes with
commensurable frequencies. This may pave a road toward large-scale
spintronic applications.

A fundamental question in this regard concerns limitations of spin
initialization and detection by optical pulses. In previous studies,
pumping and probing of spin excitations were done by short optical
pulses with durations no longer than a few picoseconds, much shorter
than the period of spin precession about the magnetic
field.\cite{PhysRevLett.94.227403,greilich06,A.Greilich07212006,A.Greilich09282007,carter:167403}
The spin initialization is most efficient for laser pulses with area
$\Theta = \pi$, requiring high peak powers generated by bulky lasers
such as Ti-Sapphire oscillators, for example, pumped by intense
continuous wave lasers.

For applications, the use of more compact pulsed solid state lasers
is appealing, which typically provide, however, considerably lower
output power levels. To reach a pulse area of $\pi$ then, the pulses
must have much longer duration. On the other hand, such pulses can
also have a much smaller spectral width, so that a less
inhomogeneous QD distribution is excited. But the condition that
these pulses are much shorter than the precession period is not
necessarily fulfilled then, potentially affecting the spin
coherence. Usage of lasers with reduced peak powers may be also
beneficial in other respects, for example the reduced importance of
non-linear optical processes such as two- or multi-photon
absorption, which may serve as potential sources of spin
decoherence.

Here we address this problem by reporting on mode-locked spin
coherence initialization and detection in $n$-type singly-charged
QDs, using pump and probe pulses with durations up to 80~ps. We
demonstrate experimentally, that the efficiency of initialization
and detection depends strongly on the ratio of laser pulse duration
to spin precession period. The experimental data are in good
agreement with predictions based on a microscopic model.

The paper is organized as follows: In Sec.~\ref{Sec:exp} the optical
techniques are described and the experimental results are presented
in Sec.~III. Sec.~\ref{Sec:model} provides the theoretical
background, and the comparison between experiment and theory is
discussed in Sec.~V.

\section{Sample and experiment}
\label{Sec:exp}

We study the spin coherence in an (In,Ga)As/GaAs self-assembled QD
ensemble, grown by molecular-beam epitaxy. The sample contains $20$
layers of (In,Ga)As dots, separated by $60$~nm GaAs barriers. The QD density
in each layer is about $10^{10}$~dots/cm$^2$. $\delta$-sheets of Si
donors are positioned $20$~nm below each QD layer with a dopant
density roughly equal to the dot density to achieve an average
occupation of one resident electron per QD. The sample was thermally
annealed at a temperature of $945$~$^\circ$C for $30$~s. It is
mounted in a superconducting split-coil magnet cryostat which allows
application of magnetic fields $B$ up to $6$~T. The sample is cooled
down to $T=6$~K by helium contact gas. At this  temperature, the ground state
photoluminescence maximum is at $1.398$~eV. For monitoring the spin
precession, an external magnetic field is applied perpendicular to
the light propagation direction (Voigt geometry).

The spin precession is traced by time-resolved pump-probe
techniques. A Ti:Sapphire laser emits pulses at a repetition rate of
$75.75$~MHz, corresponding to a repetition period $T_R=13.2$~ns. The
range of pulse durations that can be covered with this laser extends
from less than 100~fs up to 80~ps. In all cases the precise pulse
duration depends on the laser adjustment with variations on the
order of 10\%. In the case of sub-ps pulses this is not relevant for
the physics described below, but for the few 10 ps pulses this leads
to slight variations of the measured spin coherence signal, without
affecting the general conclusions. We also note, that the spectral
width of the pulses decreases inversely with the pulse duration increase, but
the pulses are not Fourier-limited for pulse durations exceeding
2~ps.

The laser beam is split into a pump beam and a probe beam, both having the
same photon energies resonant to the ground state photoluminescence
peak so that they excite the singlet trion transition. Spin
polarization of resident electrons and electron-hole complexes is
induced by the pump beam, which is modulated by a photoelastic
modulator, varying between left- and right-handed circular
polarization at a frequency of $50$~kHz. The intensity of the pump
is about $5$ times higher than that of the linearly polarized probe
beam. Independent of the pump pulse duration the laser output power
was adjusted in order to obtain maximal signal amplitude, which is
achieved by a pump pulse area of about $\Theta = \pi$.

After transmission through the sample the probe beam is split into
two orthogonal polarizations, whose intensities are detected by a
balanced photodiode bridge. Depending on the $z$-component of the
spin polarization (where $z$ is the light propagation direction) the
plane of linear polarization of the probe beam is rotated due to the 
Faraday rotation (FR) effect which leads to a variation of the intensities of the
two split beams. By polarizing the two beams appropriately, either
Faraday rotation or ellipticity is measured.\cite{glazov2010a} The time delay between
pump and probe pulses is tuned by a mechanical delay line up to $13$~ns
with a precision of about $20$~fs.

This setup is used as long as resonant pump and probe pulses of the
same duration are applied. When varying these durations relative to
each other, the setup is modified. One laser is then used as pump
(probe) only, while a second laser is used as probe (pump). The
duration of the pulses emitted from this second laser is fixed at 2~ps. 
Both lasers are synchronized with an accuracy of about 100~fs by
using one of them as master laser for the second laser, whose pulse
repetition rate is adjusted accordingly. The photon energies of the
pump and probe pulses are kept in resonance with an accuracy of
$0.1$~meV.

\section{Experimental results}
\label{Sec:results}

\begin{figure}[htbp]
\includegraphics[width=\linewidth]{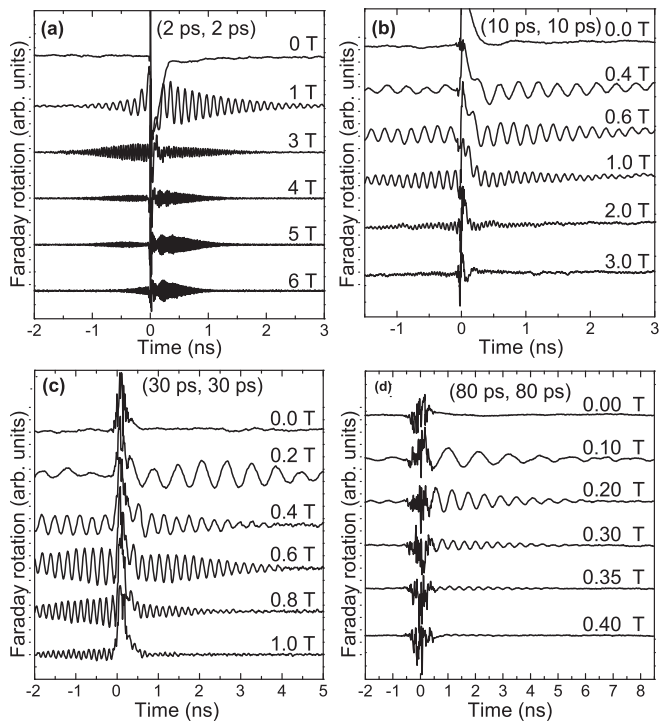}
\caption{Faraday rotation signals measured as function of time delay
between pump and probe at different magnetic fields. The equal pump
and probe pulse durations were $\tau_{\rm pump} = \tau_{\rm probe}
=2$~ps (a), $10$~ps (b), $30$~ps (c), and 80$\,$ps (d), as indicated
by the numbers in brackets giving $(\tau_{\rm pump}, \tau_{\rm
probe})$. The noise around zero delay comes from scattered laser
light, as seen particularly well for the long duration pulses.}
\label{fig:exp1}
\end{figure}

Figure \ref{fig:exp1} shows time resolved Faraday rotation signals,
where the durations of pump and probe pulses are equal, $\tau_{\rm
pump} = \tau_{\rm probe}$. Different panels correspond to different
pulse durations of 2, 10, 30, and 80~ps. In each case different magnetic field strengths $B$
are applied.

For $2$~ps pulses strong Faraday rotation signals appear up to the
highest applicable magnetic fields, so that in all cases also mode-locked
spin coherence can be generated and detected, in agreement with
previous reports.\cite{A.Greilich07212006} A prerequisite outlined
in these studies is that the pump pulse duration is much shorter
than the period of spin precession, given by $T_e = 2 \pi \hbar
/(g_e \mu_B B)$, where $\hbar$ is the Planck constant and $\mu_B$ is
the Bohr magneton. $g_e$ is the average electron $g$ factor of the
optically excited QD electron spin ensemble. For our QDs with a
$g$ factor of -0.56 at the ground state photoluminescence maximum we
obtain $T_e$ [ps] $= 127/B$[T], which gives 21 ps at $B$=6 T, still
an order of magnitude longer than the pulse duration. Therefore the
Larmor precession is of negligible influence for the processes of
both spin coherence generation and measurement as pump and probe do
not average over distinctly varying spin orientations during
precession.

We have performed also experiments for pulse durations below 1 ps,
and the appearance of the Faraday rotation traces (not shown) for
magnetic fields up to 6 T is similar to the one for 2 ps pulses, so
that generation and detection of spin coherence work efficiently also
in these cases. Our setup permits us, however, also to increase the
laser pulse duration to being comparable or even longer than
the spin precession period. For this long pulse case the question
arises to what extent the spin coherence can still be accessed.

Faraday rotation traces for $10$~ps pulses are shown in Fig.~\ref{fig:exp1}(b). The curves
recorded for low magnetic fields show strong spin precession signal,
but beyond 1~T the signal strength gets continuously weaker. Above
about 3~T spin precession can no longer be resolved. As a
characteristic quantity for this transition we use the product of
Larmor precession frequency, $\Omega_{\rm L} = 2\pi / T_e$, times the pump
pulse duration: $\Omega_{\rm L} \tau_{\rm pump}$. The field strength
of 3 T corresponds to a value of 1.5 for this product. This means
that during the pump pulse the spins perform about a quarter of a
full revolution about the magnetic field. The characteristic value
of 1.5 for this product is also found when the pulse duration is
extended further. For example, for 30~ps pulses in Fig.~\ref{fig:exp1}(c) strong
spin coherent signal can be observed up to 0.6~T, beyond the FR
signal drops strongly, so that above 1~T the signal strength reaches
the noise level. The 1~T field corresponds again to a product of
1.5, as confirmed for pump pulses of 80~ps where the spin coherent
signal appears up to 0.4~T only, while for higher fields it cannot
be observed anymore.

Note, however, that the magnetic field dependence of the Faraday rotation
signal amplitude is quite complicated, because it typically shows a
non-monotonous variation with $B$, as can be seen, for example, for
the (30~ps, 30~ps) configuration in Fig.~\ref{fig:exp1}(c). After being strong at low fields,
it drops around 0.4~T, gets stronger again around 0.6~T, and finally
vanishes at higher magnetic fields. The origin for this variation is
not fully clear yet, as several effects may become relevant such as
the variation of the number of mode-locked modes with increasing
field, which is particularly relevant at low fields, where only a
few precession modes are synchronized. In addition, the
nuclear-induced electron spin precession frequency focusing effects
involved in the mode locking may vary with field strength.\cite{A.Greilich09282007} 
Independent of that, the disappearance of signal at a characteristic
field where
 $\Omega_{\rm L}\tau_{\rm pump} \sim 1.5$ is valid for all
pulse durations.

In a nutshell, spin polarization can be excited and detected by
pulses with widely varying durations up to 80~ps. The mode-locking
of electron spin coherence is pronounced for any $\tau_{\rm pump}=
\tau_{\rm probe}$ as evidenced by the strong signal at negative
delays. The efficiency of spin coherence generation depends,
however, critically on the parameter $\Omega_{\rm L} \tau_{\rm
pump}$. The product $\Omega_{\rm L} \tau_{\rm pump}$ has to be
smaller than about 1.5, for higher values spin coherence
initialization and measurement do not work anymore. For completeness
we note that corresponding ellipticity traces look qualitatively
similar to the Faraday rotation traces, even though there are
quantitative differences concerning signal amplitudes and in
particular the ratio of signals before and after pump pulse
application.\cite{glazov2010a}

In all cases dephasing of the signal is seen on time scales of a few
nanoseconds. There are two reasons for this dephasing: the
excitation of an inhomogeneous spin ensemble with varying
$g$ factors and therefore also varying precession frequencies and
the spin precession about the randomly oriented nuclear magnetic
field. The latter is important mostly at low magnetic fields, while
the $g$ factor inhomogeneity becomes dominant for fields
exceeding by far the nuclear field of about 10~mT.\cite{Auer}

\begin{figure}[t]
\includegraphics[width=0.7\linewidth]{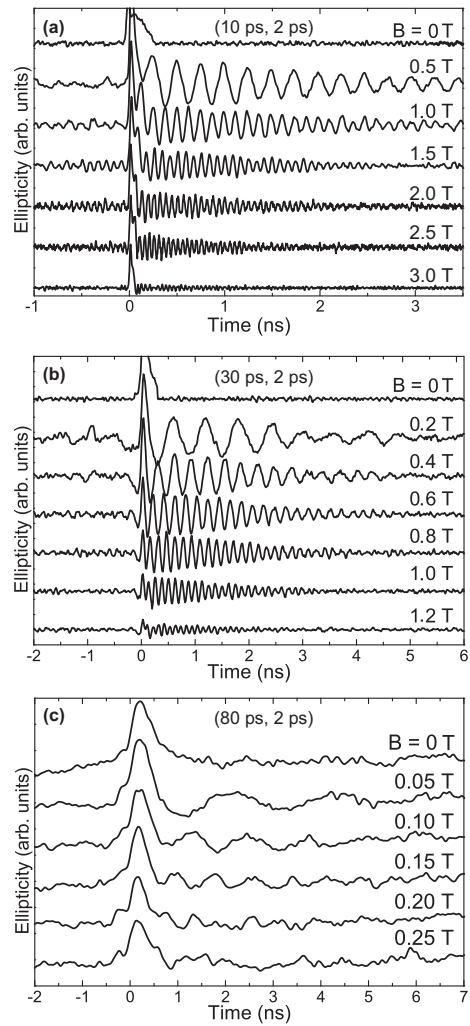}
\caption{(a) Ellipticity signals recorded with a probe pulse
duration of 2\,ps, with the pump pulse duration increased to
10~ps in panel (a), 30~ps in panel (b) and 80~ps in panel (c) at
different magnetic field strengths. The numbers in brackets give the
pump and probe durations $(\tau_{\rm pump}, \tau_{\rm probe})$.}
\label{fig:exp2}
\end{figure}

In the measurements presented so far, we use pump and probe
pulses of the same duration. For the long pulses this means that
considerable spin precession of the involved carriers, either
resident or photoexcited, occurs during pulse application. This
concerns both the initialization of coherence and also its
measurement. Ideally these two processes should be separated from
one another, which requires independent variation of pump and probe
pulse durations relative to each other $(\tau_{\rm pump} \ne
\tau_{\rm probe})$.

To that end, we first make experiments, in which the pump pulse
duration is varied, while the probe pulse duration is kept constant
at 2~ps. From above we know that this pulse duration is short enough
that the spin orientation can be considered as frozen. The results
are shown in Fig.~\ref{fig:exp2}. Figure~\ref{fig:exp2}(a) shows the magnetic field series of
ellipticity traces for 10~ps pump pulses. Note that traces measured
in Faraday rotation show a similar variation with magnetic field,
but the signal strength is weaker, as seen from the enhanced noise in the signals. As soon as $\Omega_{\rm L}
\tau_{\rm pump}$ passes a certain threshold with increasing $B$, the
coherent signal drops considerably, indicating that spin
initialization does not work anymore.

When determining the threshold magnetic field, care needs to
exercised. On the one hand, due to the longer pump pulse, having correspondingly a
reduced spectral width, a smaller number of spins becomes
initialized. As a result, a smaller number of mode-locked spin
precession modes are involved. One the other hand, the probe still
collects signals from a large, partially disordered ensemble.
Therefore the signals become overall weaker with increasing pump
duration as compared to the short pulse excitation in Fig.~\ref{fig:exp1}.

Despite of the weak signal, we see that at 3~T, which was the threshold field for the (10~ps, 10~ps) configuration, the signal gets weak, but can still be observed in Fig. 2(a). This indicates that the threshold field may be slightly higher than in the case of equal pulse duration, and that not only the pumping is influential for the signal, but also the probing. But
the difference is small as the spin initialization is hampered when the laser pulse duration exceeds a
quarter of revolution during precession. Note also that the dependence of the ellipticity signal amplitude on magnetic field is smooth and shows no strong nonmonotonic variations with $B$ as observed in Fig.~\ref{fig:exp1}. Below we will show that this dependence can be explained accounting for the finite pulse duration.

These findings are corroborated when shifting to 30~ps pump pulses
[results shown in Fig.~\ref{fig:exp2}(b)]. The magnetic field at which the signal
drop occurs is about 1.2~T, where the product $\Omega_{\rm L}
\tau_{\rm pump}$ is 1.8. This again indicates that the threshold
field is slightly larger than in the duration degenerate
configuration in Fig.~\ref{fig:exp1}. In addition, as was also the case in Fig.~\ref{fig:exp2}(a), the drop of the signal amplitude before final disappearance is
much more abrupt than in Fig.~\ref{fig:exp1}. 
Hence, the ellipticity signal
remains significant up to fields very close to the threshold field
and then drops rather fast to zero. That the coherent signals remain significant up to fields close to the threshold, in contrast to the observations in Fig.~\ref{fig:exp1}, is another indication for the importance of the probing process. 

\begin{figure}[t]
\includegraphics[width=0.7\linewidth]{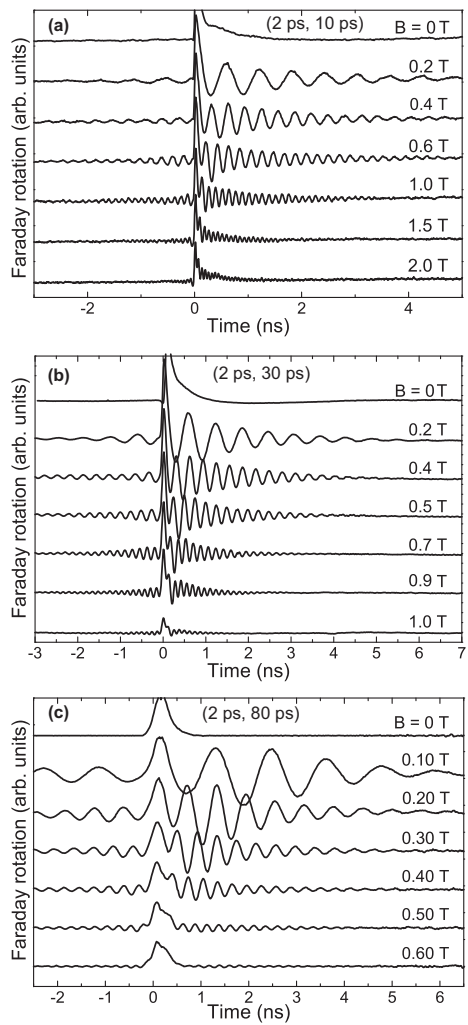}
\caption{Faraday rotation signals measured for fixed pump pulse duration
of 2 ps, but varying probe pulse durations of 10~ps (a), 30~ps (b), and 80~ps (c) at different magnetic fields. The numbers in brackets give
the pump and probe durations $(\tau_{\rm pump}, \tau_{\rm probe})$.}
\label{fig:exp3}
\end{figure}

The overall weak signal for all configurations shown in Fig.~\ref{fig:exp2}
becomes particularly pronounced for the 80~ps pump pulse case
which is presented in
Fig.~\ref{fig:exp2}(c). Here faint spin oscillations are seen at positive
delays in magnetic fields up to 0.15~T. At higher fields the noise
level exceeds the signal amplitude, so that a validation of the
threshold criterion is not possible, despite of long
accumulation times used in our experiments. 

To work out the influence of the pumping, we also test the complementary situation by fixing the
pump pulse duration at 2 ps, and changing the probe pulse duration
$\tau_{\rm probe}$. By doing so we isolate the effect of the probe
on the spin coherence measurement. The probe then has a spectral
width always smaller than the pump, so that it tests a spin ensemble
smaller than the one addressed the pump, but fully initialized.
Typical examples are shown in Fig. 3 for probe pulse durations of 10
ps (a), 30 ps (b) and 80 ps (c). In all cases strong signals are
seen, much stronger than in Fig. 2, as seen from the smooth, almost noise-free FR traces, indicating that spin initialization works well. In contrast to the 2~ps probe pulse
case where spin coherence can be detected up to 6~T, see Fig.~\ref{fig:exp1}(a), we observe here
coherent signal only up to 2~T for 10~ps probe, up to 1~T for 30~ps
probe, and up to 0.5~T for 80~ps probe. These data suggest that the
dependence of the spin signal strength on the parameter $\Omega_{\rm
L} \tau_{\rm pump}$ may be transferred also to the probe duration
dependence, for which $\Omega_{\rm L} \tau_{\rm probe}$ would be the
proper characteristic quantity.

\begin{figure}[t]
\includegraphics[width=0.9\linewidth]{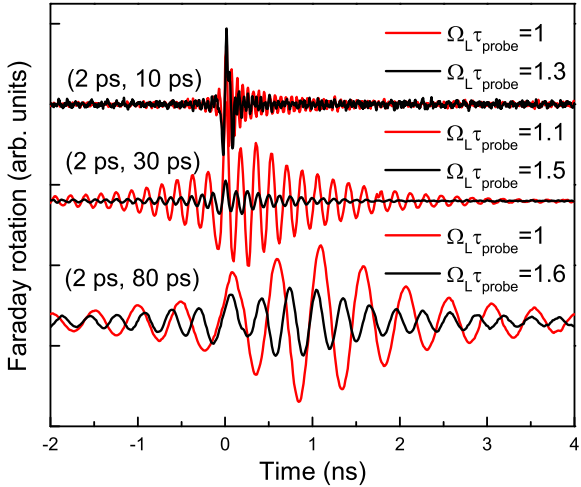}
\caption{(Color online) Faraday rotation signals measured such that
the product $\Omega_{\rm L} \tau_{\rm probe}$ is constant at a value
of 1 and 1.5 (except of the 10~ps probe case where the trace for a
product value of 1.3 is shown instead of 1.5, because at 1.5 the
signal is already very weak). For different probe durations of 10,
30, and 80~ps, the magnetic field strength is adjusted
correspondingly, as seen from the varying precession frequencies.
The pump pulse duration is fixed at 2~ps. The numbers in brackets
give pump and probe durations $(\tau_{\rm pump}, \tau_{\rm
probe})$. Magnetic fields for the curves are: (i) 2/10 ps:  $B=2$~T for $\Omega_{\rm L} \tau_{\rm
probe} =1$ and 2.5~T for $\Omega_{\rm L} \tau_{\rm
probe} =1.3$, (ii) 2/30 ps: 0.7~T for $\Omega_{\rm L} \tau_{\rm
probe} =1.1$ and 1~T for $\Omega_{\rm L} \tau_{\rm
probe} =1.5$, (iii) 2/80 ps: 0.25~T for $\Omega_{\rm L} \tau_{\rm
probe} =1$ and 0.4~T for $\Omega_{\rm L} \tau_{\rm
probe} =1.6$.} \label{fig:exp4}
\end{figure}

Also here the threshold for $\Omega_{\rm L} \tau_{\rm probe}$ is about 1.5, above which the signal drops fast to zero. Below this threshold the signal amplitude remains considerable as is validated also by Fig.~\ref{fig:exp4}, which shows Faraday rotation traces
taken for products $\Omega_{\rm L} \tau_{\rm
probe}$ equal to 1 and 1.5. For that purpose, different magnetic
field strengths are applied for the probe durations of 10, 30,
and 80~ps, as seen by the widely varying precession frequencies.
When exceeding a product value of unity in every case a
considerable drop in Faraday rotation signal strength is observed. The drop occurs, however, rather abruptly when approaching the threshold, similar as in Fig. 2 for ellipticity, but for all probe durations the magnetic field dependencies are smooth, showing no fluctuations..

All together, we find that the efficiency of spin initialization
(measurement) depends sensitively on the pump (probe) duration. The
effects of the duration increase for pump and probe have a rather symmetrical impact on the measured signal of spin
coherence.

\section{Theoretical Model}\label{Sec:model}

From the results described so far we find a strong dependence of the
spin coherence signal on both pump and probe pulse duration.
Therefore a model description of such measurements needs to take
into account generation and detection of spin coherence by finite
duration pulses. The allowance is made also for the
detuning between pump/probe pulse energies and the trion resonance in an
individual QD.

\subsection{Basic theory}\label{sec:gen}

We consider $n$-type singly-charged QDs pumped and probed by optical
pulses propagating along the sample growth axis $z$. We
assume that the optical frequencies of pump, $\omega_{\rm pump}$,
and probe, $\omega_{\rm probe}$ pulses are close to the one of the
singlet $X^-$ trion resonance with transition frequency $\omega_0$.
The QD is subject to a magnetic field which is assumed to be applied
along the $x$ axis in the dot plane. The magnetic field induces spin
splittings of the electron and trion states. The trion splitting is
neglected hereafter because the in-plane heavy-hole $g$ factor is
small as compared with the electron $g$ factor.~\cite{Mar99}

We take into account the finite durations of pump ($\tau_{\rm
pump}$) and probe ($\tau_{\rm probe}$) pulses, in contrast to
Refs.~\onlinecite{shabaev:201305,economou:205415,yugova09} where
these pulses were considered as negligibly short. In particular, we
assume that the pulse duration can be comparable or even longer than
the electron spin precession period in magnetic field, {$T_{e}$}. It
is supposed, however, that the pulses are short as compared with the
relaxation times in the system $\tau_{\rm pump}, \tau_{\rm probe}
\ll \tau_{T},\tau_{QD}$ where $\tau_{T}$ is the spin relaxation time
of the hole in trion and $\tau_{QD}$ is the trion lifetime in a
quantum dot. While the first one is at least in the 100 ns range at
$T <$ 10 K, Ref.[\cite{hsk}], the trion lifetime is 500 ps, as determined
from time-resolved photoluminescence. \footnote{In the opposite
limit the pump and probe pulses can be considered as a constant wave
radiation.} Therefore, the description of pumping and probing can be
carried out using the Schr\"{o}dinger equation without introducing a
spin density matrix.

\subsection{Generation of electron spin coherence}

We describe the QD state by a four component wave function
$\Psi=[\psi_{1/2}, \psi_{-1/2},\psi_{3/2},\psi_{-3/2}]$ where the
subscripts $\pm 1/2$ refer to the electron states and the subscripts
$\pm 3/2$ refer to the heavy-hole trion states. For a $\sigma^+$
polarized pump pulse these components obey the following equations
\begin{subequations}
\label{schroed:pump}
\begin{equation}
\mathrm i \hbar \dot\psi_{1/2} = V_+^*(t)\psi_{3/2} + \frac{\hbar\Omega_{\rm L}}{2}\psi_{-1/2},
\end{equation}
\begin{equation}
\mathrm i \hbar \dot\psi_{-1/2} = \frac{\hbar\Omega_{\rm L}}{2}\psi_{1/2},
\end{equation}
\begin{equation}
\mathrm i \hbar \dot\psi_{3/2} = \hbar\omega_0\psi_{3/2}+V_+(t)\psi_{1/2},
\end{equation}
\end{subequations}
where $V_+(t)= e^{-\mathrm i \omega_{\rm pump} t} f_{\rm
pump}(t)/\hbar$, with $f_{\rm pump}(t)$ being the smooth envelope of
the pump electric field, is the time-dependent matrix element
describing the interaction of a $\sigma^+$ polarized photon with a
QD. This matrix element is proportional to the electric field of the
pump pulse and the transition dipole matrix element.\cite{yugova09}
In the following $f_{\rm pump}(t)$ is assumed to be an even function
of time with the maximum at $t=0$. This time moment coincides with
the pump pulse arrival. Accounting for the Zeeman splitting in
Eqs.~\eqref{schroed:pump} is a major difference between the present
approach and previous
treatments.\cite{shabaev:201305,economou:205415,yugova09} Note, that
spin pumping of a free, two-dimensional gas by pulses long as
compared with the spin precession period was considered in
Ref.~\onlinecite{averkiev08}.

Without optical pumping, $V_+(t)\equiv 0$, the spin
system~\eqref{schroed:pump} precesses coherently about the in-plane
magnetic field. It is convenient to introduce the electron spin
state combinations
\begin{equation}
\label{x:spins}
\psi_{x} = \frac{1}{\sqrt{2}}(\psi_{1/2}+\psi_{-1/2}),\quad \psi_{\bar x} = \frac{1}{\sqrt{2}}(\psi_{1/2}-\psi_{-1/2}),
\end{equation}
that correspond to the eigenstates in magnetic field $\bm B
\parallel x$ and evolve in time as
\begin{equation}
\label{precess}
\psi_{x}(t) = \tilde{\psi}_x\ {\exp{({-}\mathrm i \Omega_{\rm L} t/2)}}, \quad \psi_{\bar x}(t) = \tilde{\psi}_{\bar x}\ {\exp{(\mathrm i \Omega_{\rm L} t/2)}},
\end{equation}
where $\tilde{\psi}_{x}$, $\tilde{\psi}_{\bar x}$ are constants
determined by the initial conditions. With optical pumping, $V_+(t)
\ne 0$, and $\tilde{\psi}_{x}$, $\tilde{\psi}_{\bar x}$ become
time-dependent.\cite{ll3_eng}

In order to describe the pump action on the electron spin we have to
establish a link between the spin components before and after  pump
pulse arrival. The change of spin with time results from two
effects: the Larmor precession about the magnetic field and the
effect of the pump pulse. The Larmor precession of the electron spin
during time $T$ can be described by a linear operator $\mathcal
R_{\Omega}(T)$.\cite{merkulov02} It is convenient to treat the spin
precession separately from the optical pulse and connect the rotated
electron spin vector $\bm S^{-} = \mathcal R_{\Omega}(T_0) \bm
S(-T_0)$ with the electron spin vector $\bm S^{+} = \mathcal
R_{\Omega}^{-1}(T_0) \bm S(T_0)$, where $T_0$ exceeds by far the
pulse duration so that on the time scale of $T_0$ the pulse action
can be neglected. On the quantum-mechanical level this operation is
equivalent to the unitary transformation Eq.~\eqref{precess}. Hence
the components of the spin vector $\bm S^{\pm}$ are given by:
\begin{eqnarray}
\label{spin}
S_x^{\pm} &= &\frac{1}{2}\left[|\tilde{\psi}_{x}(\pm \infty)|^2 - |\tilde{\psi}_{\bar x}(\pm \infty)|^2\right],\nonumber \\
S_y^{\pm} &=& -\Im\{\tilde{\psi}_{x}(\pm \infty)\tilde{\psi}^*_{\bar x}(\pm \infty)\},\\
S_z^{\pm} &=& \Re\{\tilde{\psi}_{x}(\pm \infty)\tilde{\psi}^*_{\bar x}(\pm \infty)\}. \nonumber
\end{eqnarray}

\begin{widetext}
One can show that in the limit of low pump power (see
Appendix~\ref{app:transform} for details)
\begin{subequations}
\label{spin:transf}
\begin{multline}
\label{Sz} S_z^{+} =-\frac{1}{2} \Re G(\Lambda,\Omega_{\rm L}) +
\Re{\left[1 -\frac{1}{2} G\left(\Lambda+\frac{\Omega_{\rm
L}}{2},0\right) -\frac{1}{2} G\left(\Lambda-\frac{\Omega_{\rm
L}}{2},0\right)\right]}S_{z}^{-} +\\ \frac{1}{2} \Im {\left[
G\left(\Lambda+\frac{\Omega_{\rm L}}{2},0\right) +
G\left(\Lambda-\frac{\Omega_{\rm L}}{2},0\right)\right]}S_{y}^{-},
\end{multline}
\begin{multline}
\label{Sx} S_x^{+} =-\frac{1}{4} \Re{\left[
G\left(\Lambda+\frac{\Omega_{\rm L}}{2},0\right) -
G\left(\Lambda-\frac{\Omega_{\rm L}}{2},0\right)\right]} +
\Re{\left[1 -\frac{1}{2} G\left(\Lambda+\frac{\Omega_{\rm
L}}{2},0\right) -\frac{1}{2} G\left(\Lambda-\frac{\Omega_{\rm
L}}{2},0\right)\right]}S_{x}^{-} - \\  \Im {G(\Lambda,\Omega_{\rm
L})}S_{y}^{-},
\end{multline}
\begin{multline}
\label{Sy} S_y^{+} =
\Re{\left[1-\frac{1}{2}G\left(\Lambda+\frac{\Omega_{\rm
L}}{2},0\right)-\frac{1}{2} G\left(\Lambda-\frac{\Omega_{\rm
L}}{2},0\right)\right]}S_{y}^{-} +  \Im {G(\Lambda,\Omega_{\rm
L})}S_{x}^{-}-\\  \frac{1}{2} \Im {\left[
G\left(\Lambda+\frac{\Omega_{\rm L}}{2},0\right) +
G\left(\Lambda-\frac{\Omega_{\rm L}}{2},0\right)\right]}S_{z}^{-}.
\end{multline}
\end{subequations}
where $\Lambda=\omega_{\rm pump} - \omega_0$ is the energy detuning
between pump pulse and trion resonance. Here
\begin{equation}
\label{g:func} G_{\rm pump}(\Lambda, \Omega) =
\int_{-\infty}^{\infty} \mathrm dt f_{\rm pump}(t) \int_{-\infty}^t
\mathrm dt' f_{\rm pump}(t') \mathrm \ e^{\mathrm i \Lambda (t-t')}
\cos{\left[\frac{\Omega}{2}(t+t')\right]}.
\end{equation}
The function $G_{\rm pump}(\Lambda, \Omega)$ can be found
analytically for Fourier-limited pulses, such as in the case of an
exponential pulse, $f_{\rm P}(t) =f_0 \mathrm e^{-|t|/\tau_{\rm
pump}}$, where $f_0$ is the pump pulse amplitude that is related to
its area, $\Theta = 2\int\limits_{-\infty}^\infty f_{\rm pump}(t)
\mathrm d t$, by $f_0=\Theta/(4\tau_{\rm pump})$. Then one can show
that
\begin{equation}
\label{g:res} G_{\rm pump}(\Lambda,\Omega) =
\frac{\Theta^2(2+\mathrm i \Lambda\tau_{\rm
pump})}{[4+(\Omega\tau_{\rm pump})^2][4-8\mathrm i \Lambda\tau_{\rm
pump} -4(\Lambda\tau_{\rm pump})^2 + (\Omega\tau_{\rm pump})^2]}.
\end{equation}
For $\Omega \tau_{\rm pump}=0$, Eq.~\eqref{g:res} is equivalent to
Eq.~(61) of Ref.~\onlinecite{yugova09}.
\end{widetext}

Although the Eqs.~\eqref{spin:transf} are quite bulky, they allow
one to identify all essential physics features caused by the pump
pulse application. First of all, circularly polarized pump pulses
cause electron spin orientation along the $z$ axis
\begin{equation}
\label{Sz+} S_z^+=-\Re{G(\Lambda,\Omega_{\rm L})/2}.
\end{equation}
In absence of a magnetic field and for resonant pulse $S_z^+ = -\Theta^2/16$. This is
equivalent to the regular spin initialization protocol based on very
short pump pulses.\cite{shabaev:201305,greilich06} Due to the Zeeman
splitting the spin can acquire some degree of orientation along the
$x$ axis due to unequal transition rates out of the magnetic field
split sublevels, see first term in Eq.~\eqref{Sx}
\begin{equation}
\label{Sx+} S_x^+=-\frac{1}{4} \Re{\left[
G\left(\Lambda+\frac{\Omega_{\rm L}}{2},0\right) -
G\left(\Lambda-\frac{\Omega_{\rm L}}{2},0\right)\right]}.
\end{equation}
This effect arises from detuning of the pump pulse from the center
of the Zeeman-split doublet, as discussed in more detail below. In
addition, the spin is rotated due to the pump pulse action,
in the $(xy)$ plane similar to the case of negligible Zeeman
splitting\cite{economou:205415,yugova09,phelps:237402} and in the
$(yz)$ plane due to the combined action of the pump pulse and the
spin splitting.

\begin{figure}[hptb]
\includegraphics[width=0.45\textwidth]{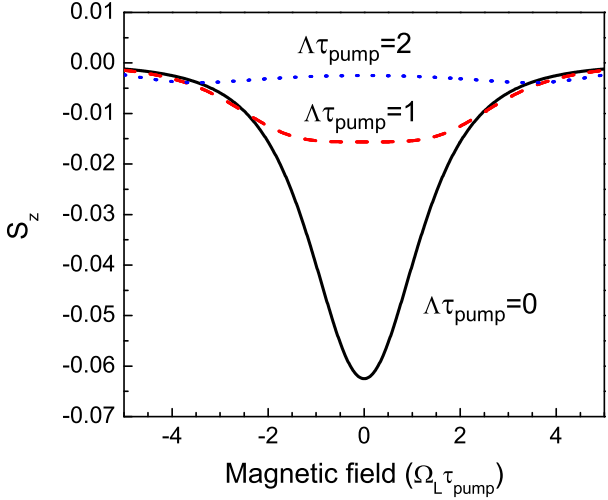}%\\
\caption{%Top panel:
(Color online) Electron spin $z$ component generated by a single
pump pulse as a function of reduced magnetic field $\Omega_{\rm L}
\tau_{\rm pump}$. The three curves correspond to different detunings
between the pump and trion resonance $\Lambda\tau_{\rm pump}=0$
(black/solid), $\Lambda\tau_{\rm pump}=1$ (red/dashed), and $\Lambda\tau_{\rm
pump}=2$ (blue/dotted). Pulse area $\Theta=1$.}\label{fig:init}
\end{figure}

\begin{figure}[hptb]
\includegraphics[width=0.45\textwidth]{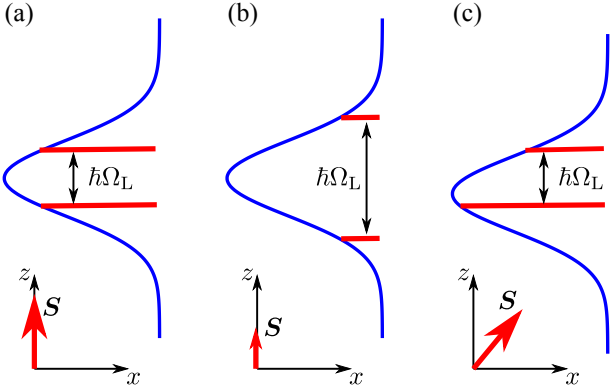}
\caption{Schematic illustration of electron spin initialization. Red
lines indicate Zeeman-split electron sublevels with splitting $\hbar\Omega_{\rm L}$, the blue curve shows
the pump pulse spectral shape. (a) and (b) give the case of a pump
pulse in resonance with the center of the Zeeman doublet. (a)
$\Omega_{\rm L}\tau_{\rm pump} \ll 1$, both sublevels interact
strongly with the optical pulse resulting in efficient spin
orientation along the $z$ axis. (b) $\Omega_{\rm L}\tau_{\rm pump}
\gg 1$, inefficient spin orientation along the $z$ axis. (c) Case of
a detuned pulse resulting in different interaction strengths with
the Zeeman-split sublevels. As a result the electron spin acquires
non-zero $x$ and $z$ components.}\label{fig:scheme}
\end{figure}

It is worth to note that if the electron spin was initially
unpolarized, $\bm S^-=0$, a circularly polarized pump pulse creates
electron spin with two non-zero components, $S_z^+$ and $S_x^+$. The
appearance of $S_z^+$ is related to the transfer of photon angular
momentum to the electron, with an efficiency that decreases with
increasing magnetic field, see Fig.~\ref{fig:init}. Indeed, with
increasing spin splitting the formation of a coherent superposition
of the Zeeman split sublevels becomes hindered as schematically
illustrated in Figs.~\ref{fig:scheme}(a) and \ref{fig:scheme}(b). The contained information can be also translated into the time domain: For a fixed Zeeman splitting, the action of a longer pulse corresponds to applying a spectrally narrower pulse, equivalent to going from (a) to (b), and reducing thereby the spin initialization efficiency.

The microscopic origin of the appearance of the in-plane spin
component $S_x^+$ is shown in Fig.~\ref{fig:scheme}(c). Indeed, if
the pump pulse is detuned from the ``center-of-gravity'' of the
Zeeman-split doublet, the transition efficiencies from the split
levels are different. As a result, the resident carrier acquires
some spin polarization parallel or anti-parallel to the magnetic
field.

As it was assumed that the trion lifetime, $\tau_{QD}$, exceeds by
far the pump duration, the dynamics of the coupled electron and
trion spins is described in the standard way, see
Ref.~\onlinecite{yugova09}, Eq. (27), and Ref.~\onlinecite{zhu07},
Eq. (6). Long living electron spin coherence appears after the trion
recombination, and a steady state distribution of the precessing
spins develops as result of the applied pump
pulses.\cite{A.Greilich07212006,yugova09}

\subsection{Detection of electron spin coherence}

The description of spin coherence probing is rather similar to the
generation. We assume that the probe is linearly polarized along the
$x$-axis, i.e. along the magnetic field direction as in our
experiment. In this case, the coupled Schr\"{o}dinger equations
describing the dynamics of electron and trion spins separate into
two independent subsystems, corresponding to the optical transitions
involving electrons with spin parallel to the $x$ axis and those
with spin antiparallel to the $x$ axis.

% The Schr\"{o}dinger equations describing the coupled dynamics of electron $\pm 1/2$ states and trion $\pm 3/2$ states can be split into two independent systems:
% \begin{subequations}
% \label{schroed:probe1}
% \begin{equation}
% \mathrm i \hbar \dot\psi_{x} = V^*(t)\psi_{t} + \frac{\hbar\Omega_{\rm L}}{2}\psi_{x},
% \end{equation}
% \begin{equation}
% \mathrm i \hbar \dot\psi_{t} = \hbar\omega_0\psi_{t}+V(t)\psi_{x},
% \end{equation}
% \end{subequations}
% and
% \begin{subequations}
% \label{schroed:probe2}
% \begin{equation}
% \mathrm i \hbar \dot\psi_{\bar x} = V^*(t)\psi_{\bar t} - \frac{\hbar\Omega_{\rm L}}{2}\psi_{\bar x},
% \end{equation}
% \begin{equation}
% \mathrm i \hbar \dot\psi_{\bar t} = \hbar\omega_0\psi_{\bar t}+V(t)\psi_{x},
% \end{equation}
% \end{subequations}
% where we introduced linear combinations $\psi_t = (\psi_{3/2}+\psi_{-3/2})/\sqrt{2}$ and $\psi_{\bar t} = (\psi_{3/2}-\psi_{-3/2})/\sqrt{2}$. Here $V(t) = e^{-\mathrm i \omega_{\rm pr} t}f_{\rm pr}/\hbar$, with $\omega_{\rm pr}$ is the probe optical frequency, is the transition matrix element induced by the linearly polarized light and $f_{\rm pr}$ is the probe envelope function.  The separation of the Scr\"{o}dinger equation into two independent subsystem is a consequence of the considered geometry where the magnetic field axis and the linear polarization plane of the probe are parallel. 

It can be shown that, similarly to
Refs.~\onlinecite{yugova09,glazov2010a}, the spin ellipticity, $\mathcal
E$, and Faraday rotation, $\mathcal F$, signals from an ensemble of
QD spins are proportional to the real and imaginary parts of the
following quantity:
\begin{equation}
\label{Sigma} \mathcal E(t) + \mathrm i \mathcal F(t) \propto G_{\rm
probe}(\omega_{\rm probe} - \omega_0,\Omega)S_z(t),
\end{equation}
where $G_{\rm probe}$ is defined by Eq.~(\ref{g:func}) after
replacing the pump envelope $f_{\rm pump}(t)$ by the probe envelope
$f_{\rm probe}(t)$. Here $S_z(t)$ is the electron spin $z$ component
at the moment of probe pulse action, i.e., at the time where the
probe pulse amplitude is maximal. In deriving Eq.~\eqref{Sigma} we
assumed that the pump-probe delay exceeds the trion spin lifetime in
the QD, $\tau_T\tau_{QD}/(\tau_T+\tau_{QD})$, which makes possible
to neglect the contribution from the trion spin polarization to the
measured signal. Otherwise, an additional contribution to Eq.~\eqref{Sigma}
should be taken into account which is proportional to the
hole-in-trion spin polarization~\cite{yugova09}.

\begin{figure}[hptb]
\includegraphics[width=0.4\textwidth]{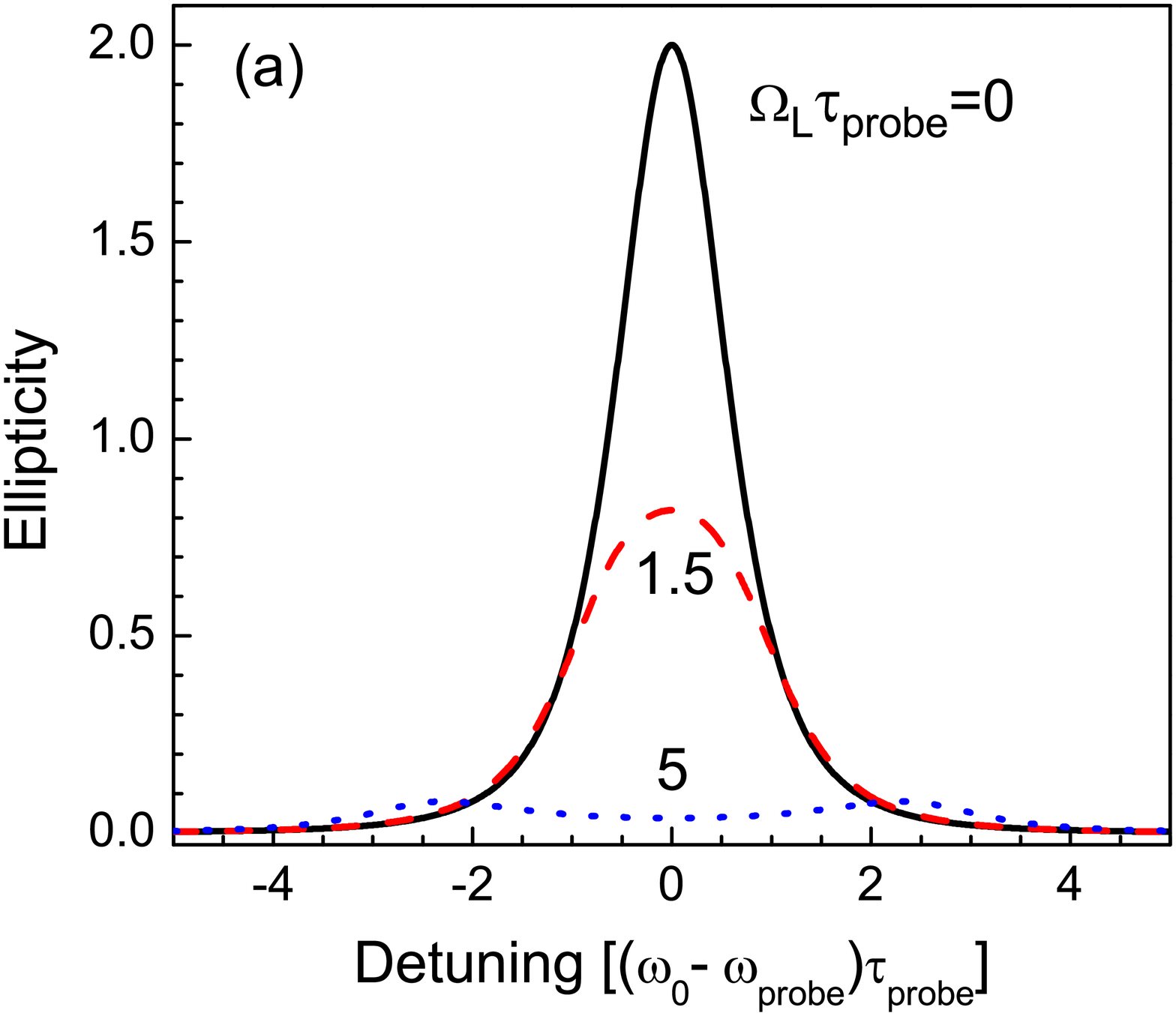}\\
\includegraphics[width=0.4\textwidth]{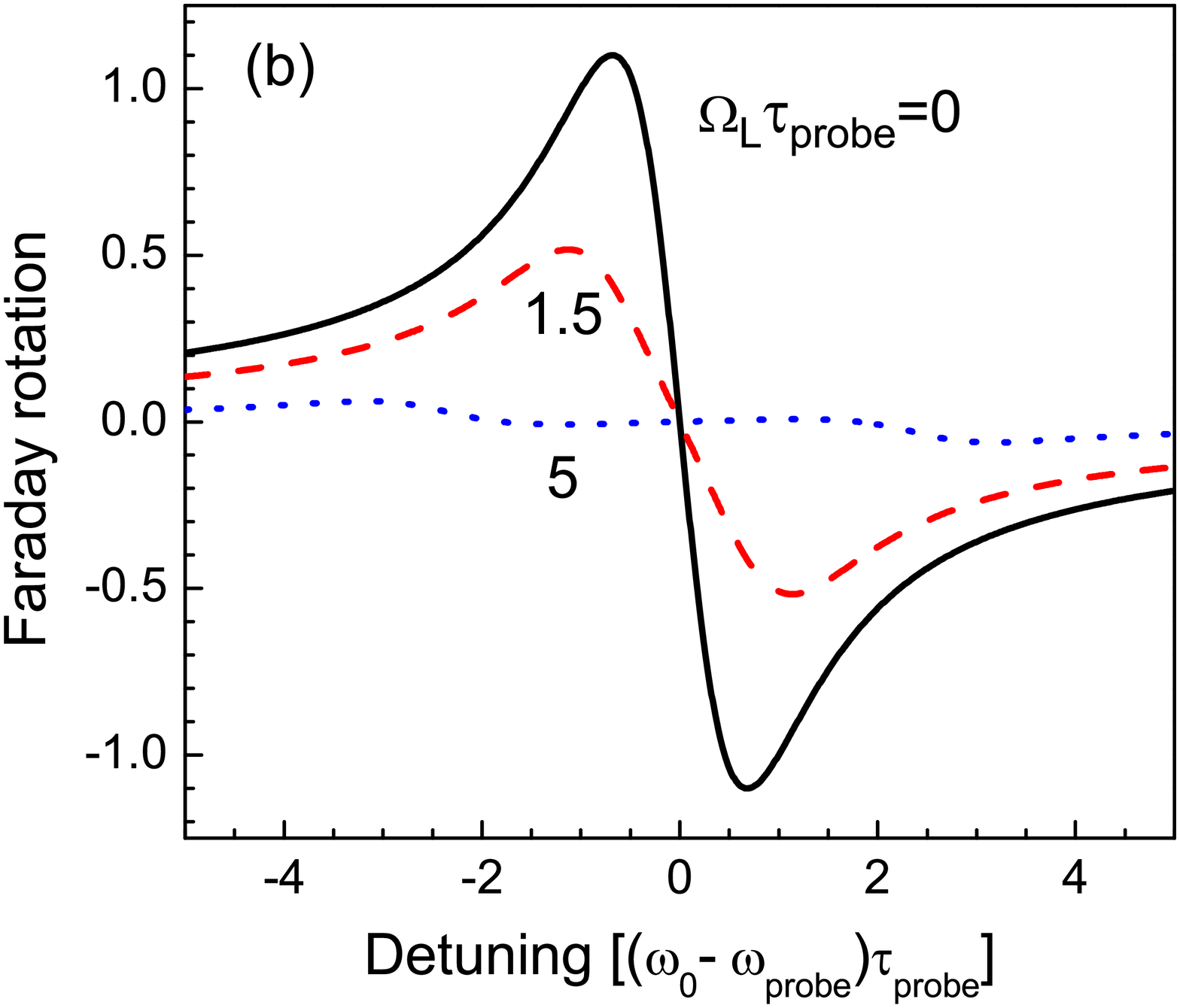}
\caption{Ellipticity (a) and Faraday rotation (b) signals as
function of detuning between the quantum dot resonance and the probe
optical frequency. The three curves correspond to different values
of magnetic field, $\Omega_{\rm L}\tau_{\rm probe}=0$ (black/solid),
$\Omega_{\rm L}\tau_{\rm probe}=1.5$ (red/dashed), and $\Omega_{\rm
L}\tau_{\rm probe}=5$ (blue/dotted). The spin $z$ component is the same for all curves. The signals are given in arbitrary
units.}
\label{fig:signals}
\end{figure}

Figure~\ref{fig:signals} shows the dependence of the ellipticty and
Faraday rotation signals on the detuning between the trion resonance
and the probe optical frequency. Note, that these signals
are calculated for a given QD, no averaging over the
ensemble is done. The overall behavior is similar to
the one known for probing by short pulses, $\Omega_{\rm L}\tau_{\rm
probe} \ll 1$ (shown by the black curves in
Fig.~\ref{fig:signals}).\cite{yugova09,carter:167403} The
ellipticity is maximal for degenerate probe and trion resonance, while the
Faraday rotation has a zero for $\omega_0 = \omega_{\rm probe}$.
With increasing magnetic field the signal strength drops
strongly, both in ellipticity and in Faraday rotation for
almost all values of the detuning.
The maximum of ellipticity transforms
into a minimum and a fine structure appears for $\Omega_{\rm
L}\tau_{\rm probe}=5$ which corresponds to the
probe tuned to the two Zeeman-split sublevels. This fine structure
becomes also visible in the Faraday rotation signal. In addition the
spectral shape of the signal changes.

The QD ensemble is inhomogeneous, and the pump pulse excites a
subensemble of dots with various trion resonance
frequencies.\cite{glazov2010a} Hence, the observed Faraday rotation
signal should be averaged over the spin distribution. The result of
this averaging depends strongly on the possible asymmetry of the
spin distribution as well as on the details of spin coherence
excitation and mode-locking. Below, we show that even the simplest
model, where the inhomogeneity is ignored and the asymmetry of the
quantum dot distribution is modeled as an effective detuning,
describes well the experimental findings.

\section{Comparison of theory and experiment}

\begin{figure}[t]
\includegraphics[width=1.1\linewidth]{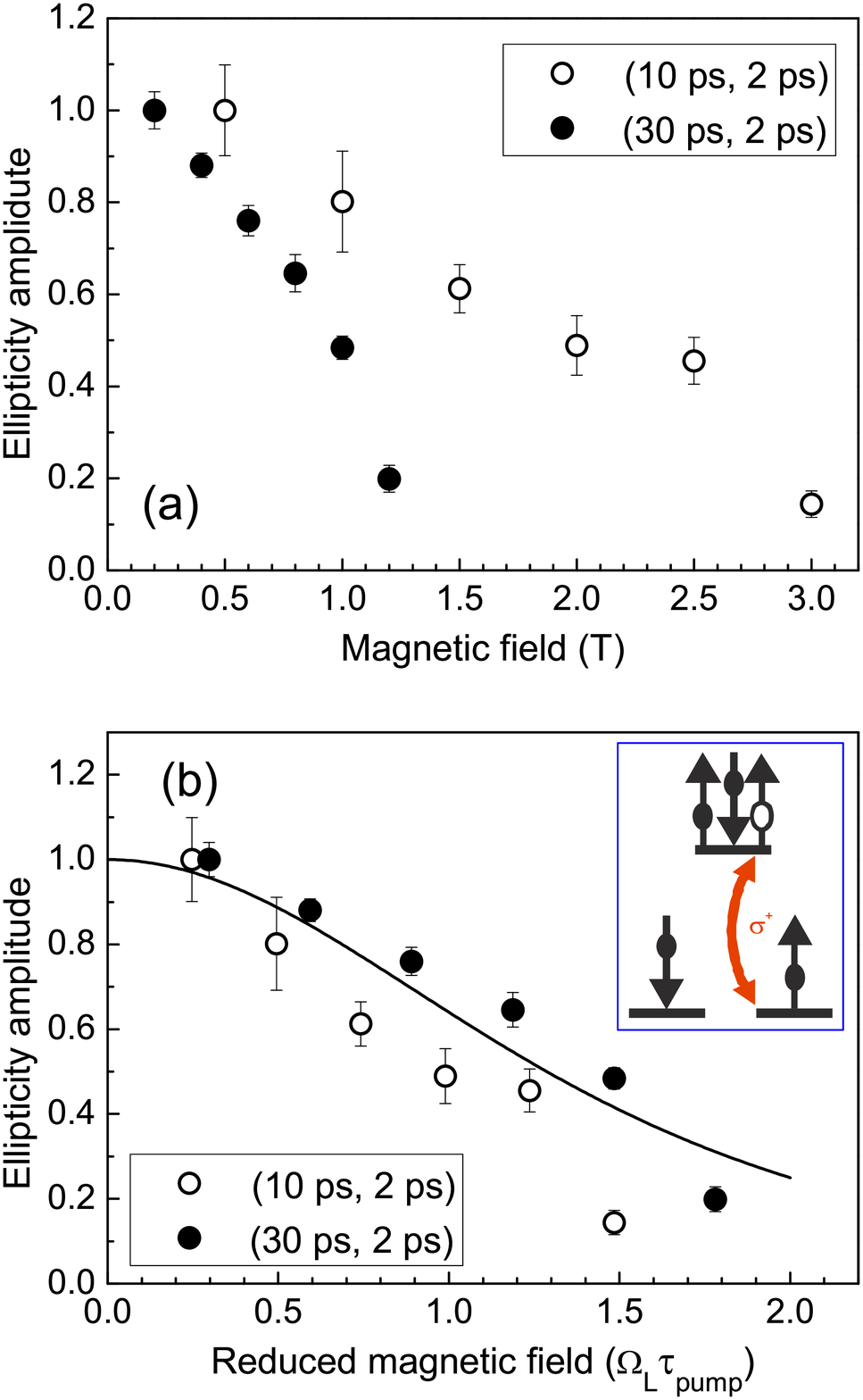}
\caption{(Color online) Ellipticity amplitude versus magnetic field
for pump pulses with a duration of 10 ps and 30 ps, probed by 2 ps
pulses. Panel (a) and (b) show the signal amplitude as function of
magnetic field and reduced field $\Omega_{\rm L}\tau_{\rm pump}$,
respectively. The solid curve in panel (b) is normalized to the
$\Omega_{\rm L}\tau_{\rm pump} =0$ dependence of $S_z^+$, which is the spin component amplitude right after pump pulse arrival, calculated according to Eq.~\eqref{Sz+}. $\omega_{\rm pump}= \omega_{\rm probe}$.} \label{fig:exp2:1}
\end{figure}

With this general theoretical setting we can compare the calculated
magnetic field dependencies of the spin coherence signal with the
measured data for different durations of pump pulse and probe pulse.
To distinguish between the effects of pump and probe, we focus first
on the experiments where one of the pulses was made longer compared
to the other pulse with duration fixed at 2~ps. The corresponding
experimental data have been shown in Figs.~2 and 3, respectively.

Here we need to comment on the shape of the pump-probe traces. In
Ref.~\onlinecite{glazov2010a}, the ellipticity signal was shown to
drop smoothly to zero when moving from zero towards negative or
positive delays. For Faraday rotation, however, the signal may rise
first before a signal drop due to dephasing is seen, when using
resonant pump and probe pulses of the same duration. The Faraday
rotation behavior described in Ref.~\onlinecite{glazov2010a} is
observed in Fig.~1, where pump and probe pulses of the same duration
were taken from a single laser. The signal rise for short delays is
particularly pronounced at 0.2~T in the 30~ps case. When detuning
pump and probe spectrally the behavior goes back to the conventional
one, like in ellipticity with maximum signal at zero delay. While
the traces in Fig.~2 show ellipticity anyway, the traces in Fig.~3
give Faraday rotation signals which, however, have a smooth drop
when moving away from time zero, despite of the targeted pump and
probe energy resonance. This makes determination of amplitudes
{quite} simple. We attribute this behavior for different pump and
probe durations to an effective detuning of the pulses arising from
their different spectral widths, where the spectral components
outside of the profile of the other laser lead to the effective
detuning. In addition the accuracy of putting the pulses from the
two lasers in resonance was about 0.1~meV, potentially leading to
another small detuning.

From these data we extract the spin coherence signal amplitudes, as
measure for the efficiency of either coherence generation or
readout. Focusing on the generation process, we have first done this
for pump durations of 10~ps and 30~ps, with the probe duration fixed
at 2~ps. For determining the amplitudes, the Faraday rotation traces
were fitted by exponentially damped harmonics for delay times, when
all optically excited exciton complexes have decayed. The amplitudes
for different magnetic fields were then normalized by the amplitude
for the magnetic field, where the amplitude was maximum.

We also note here, that in contrast to the theoretical modeling (see
below) maximum amplitude is reached for finite magnetic fields. We
attribute this to the effects of spin precession of the hole in the
optically excited trion: the hole $g$ factor $g_h$ has been measured
to be small, but non-zero ($g_h \approx 0.12$) in the structure
under study. In low magnetic fields electron spin coherence appears
due to the hole spin relaxation only which results in depolarization
of the electron left after trion recombination. In higher fields the
hole precesses during the trion lifetime which leads in effect to
its spin relaxation. As a result, the long-lived electron spin
coherence increases in the range of small fields \cite{sokolova}. In addition, at
low external fields nuclear effects can come into play leading to
electron depolarization. Therefore we expect a maximum of the
electron spin coherence signal at a finite field, $B_{max}$.
Experimentally this field lies in the range from 0.1 to 0.2~T.

The magnetic field dependence of the normalized ellipticity amplitudes for 10~ps
and 30~ps pump pulses is shown in Fig. 8(a). The amplitude for 30 ps pulses drop smoothly to
zero with increasing magnetic field up to slightly more than 1~T.
For 10~ps pump pulses the amplitude drop with increasing $B$ does
not occur as fast, but takes place over an extended field range up
to 3~T. As pointed out, this behavior can be characterized by the
reduced magnetic field product $\Omega_{\rm L} \tau_{\rm pump}$, for which we had found that the spin coherence signal basically disappears when the value of 1.5 is exceeded, as confirmed by Fig. 8(b),
showing the amplitude data as function of $\Omega_{\rm L} \tau_{\rm
pump}$. In this representation the data for the two pump pulse
durations basically coincide and converge to zero for $\Omega_{\rm L}
\tau_{\rm pump} = 1.5$, corroborating the universality of the
threshold.

The presence of a threshold, outlined already in the theory
subsection, can be qualitatively understood in terms of pulse
duration. As schematically shown in the inset of Fig.~8(b), the
trion formation is spin selective. For $\sigma^+$ polarized light
the spin-up electron contributes to the trion and gets depolarized
afterwards, while the spin-down electron does not participate in
trion formation. These are the electrons whose spin is accumulated
due to the train of pump pulses. However, if during the pump pulse
action this electron spin component has time to rotate significantly, it also participates in trion formation and becomes depolarized. Thereby the pumping efficiency is diminished.

The solid curve in Fig.~8(b) shows the theoretical result for
$S_z^+$ as a function of reduced magnetic field calculated after
Eq.~\eqref{Sz+}. The electron spin $z$ component value at the moment of pump pulse arrival is normalized by its value at $\Omega_{\rm L}\tau_{\rm pump}
=0$. The theoretical curve follows rather well the experimental
points. The disappearance of spin coherent signal $\Omega_{\rm L}\tau_{\rm pump} =1.5$ can be seen from Fig. 5: compared to the zero value for the product, the initialized spin component is reduced by a factor of about 3 for $\Omega_{\rm L}\tau_{\rm pump} = 1$ and it basically has vanished for $\Omega_{\rm L}\tau_{\rm pump} = 2$, as the threshold value of 1.5 has been crossed.

The theoretical modeling was done for Fourier-limited pulses, for which we find good agreement with the data, even though the spectral width of the pulses is somewhat larger than expected from their duration. Discrepancies between experiment and theory, also in Fig. 9 might arise, however, from this difference, or from the higher pump excitation power ($\Theta = \pi$) than assumed in theory for the pump. 

\begin{figure}[t]
\includegraphics[width=1.0\linewidth]{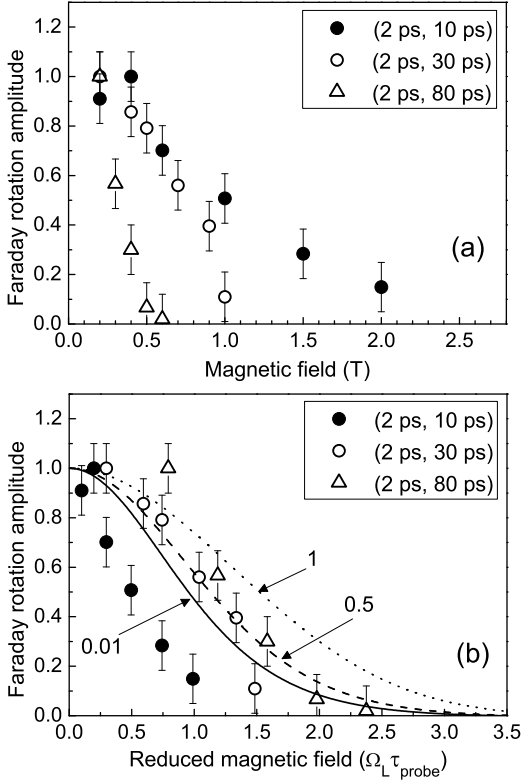}
\caption{(a) Normalized Faraday rotation amplitudes  as functions of
magnetic field: squares correspond to $\tau_{\rm probe}=10$~ps,
circles to $\tau_{\rm probe}=30$~ps and triangles $\tau_{\rm
probe}=80$~ps, while the pump pulse duration was fixed at 2\,ps. (b)
Normalized Faraday rotation amplitudes versus reduced magnetic field
$\Omega_{\rm L}\tau_{\rm pr}$ for the data taken from panel (a).
Solid, dashed and dotted curves give calculations using
Eq.~\eqref{Sigma} for different detunings between the probe pulse
and the trion resonance $(\omega_{\rm probe} - \omega_0)\tau_{\rm
  probe}=0.01$ (solid), $0.5$ (dashed) and $1$ (dotted).}\label{fig9}
\end{figure}

A similar threshold effect was observed for the influence of the
probe pulse, as can be seen from the magnetic field dependence of the Faraday rotation amplitude for different probe
durations, while the pump duration was fixed at 2~ps presented in Fig.~\ref{fig9}. The probe can
reflect the initialization of the spins by the pump only as long as
these show a dominant preferential orientation. This is the case for
probes shorter than a quarter of revolution during precession.
Therefore the drop of the Faraday signal amplitude in Fig.~9(a) occurs at higher magnetic fields when the probe pulses are shorter.

Also for this situation a kind of universal behavior is found in
which the absolute magnetic field strength is not decisive but
rather the reduced magnetic field strength $\Omega_{\rm L} \tau_{\rm
probe}$. Figure 9(b) gives the corresponding dependence of FR signal
amplitude. Considering also the experimental accuracy the data for
the three different probe pulse durations are close to being
identical, indicating a universal behavior on reduced magnetic
field. This universal behavior is in accord with the calculations
shown by lines in Fig.~9(b). The different calculated curves give
different detunings between the probe pulse and trion resonance:
$(\omega_{\rm probe} - \omega_0)\tau_{\rm probe}=0.01$ (solid),
$(\omega_{\rm probe} - \omega_0)\tau_{\rm probe}=0.5$ (dashed) and
$(\omega_{\rm probe} - \omega_0)\tau_{\rm probe}=1$ (dotted), as is
the case also in experiment. The difference between these curves is,
however, small. 

The drop of probed signal amplitude with increasing magnetic field is about the same for Faraday rotation and allipticity as can be seen From Fig. 7. Up to $\Omega_{\rm
L} \tau_{\rm probe} = 1.5$ the signal amplitude drops by a factor of 2.5 compared with the limit of short
pulses. Beyong this threshold the signal drop ocurs rather abruptly. 
The agreement of the experimental results and
theoretical calculations in the simplified model demonstrates that
the basic mechanisms of the spin coherence excitation and detection
are well understood.

The impact of the finite pump and probe durations is brought
together in the experiments with identical pump and probe durations.
However, a comparison of the absolute amplitude values is basically
impossible, as the corresponding experiments involve different
experimental schemes with either a single or two lasers with
different focusing on different sample positions in the different
measurement runs. Also the magnitude of the pumped and probed spin
ensembles varies for the different configurations which involved
pulses with either the same or significantly different spectral
widths. In addition, we found in Fig.~1 a non-monotonic dependence
of the Faraday signal amplitude on magnetic field. Therefore we do
not attempt to make a quantitative comparison. Still, qualitatively
the picture is quite transparent.

The experiments with either the pump duration or the probe duration
varied demonstrate that there is a threshold field above which spin
coherence is not found, see Figs.~\ref{fig:exp2:1}(b) and \ref{fig9}(b). In both cases the threshold field is
similar for the same pump or probe durations. As the effects of pump
and probe enter the spin coherence signal rather "symmetrically",
a prolongation of both pulses in the duration-degenerate
configuration leads to a disappearance of signal amplitude at
basically the same field strength. From the calculations we find that the spin coherence drop due to the combined action of a pump and a probe elongation lead sot a signal drop by about an order of magnitude for $\Omega_{\rm L} \tau_{\rm pump} = 1.5$ and $\tau_{\rm pump} = \tau_{\rm probe}$. This explains the disappearance of the spin coherent signal at this threshold value. 

\section{Conclusions}

To conclude, we have demonstrated theoretically and experimentally
the feasibility to initialize and detect electron spin coherence by
long optical pulses with durations up to 80~ps, comparable with the
electron spin precession period. The efficiency of electron spin
coherence measurement is determined by the ratio of the periods for
pulse duration and spin precession and with an increase of the
magnetic field the spin signals decrease. The experimental results
and theoretical calculations are in good agreement.
Based on this demonstration spin initialization by compact pulsed solid state lasers with limited output power becomes feasible in low magnetic fields, to which applications would be limited anyway in applications. 

\acknowledgments This work was supported by the Deutsche
Forschungsgemeinschaft, the Bundesmninsterium f\"ur Bildung und
Forschung project ``QuaHL-Rep, Russian Foundation of Basic
Research, ``Dynasty'' Foundation---ICFPM and EU FP7 project
Spinoptronics.

\appendix

\begin{widetext}

\section{Spin pumping by long pulses}\label{app:transform}

The pump pulse effect is conveniently described by an operator $\mathcal Q$
which transforms the pair $[\tilde \psi_{x}(-\infty)$,
$\tilde{\psi}_{\bar x}(-\infty)]$ components of the wave function
(long before the pulse arrival) into the pair $[\tilde
\psi_{x}(+\infty)$, $\tilde{\psi}_{\bar x}(+\infty)]$ (after pulse
arrival):
\begin{eqnarray}
\tilde \psi_{x}(+\infty) = \mathcal Q_{xx}\tilde{\psi}_x(-\infty) +\mathcal Q_{x\bar x}\tilde{ \psi}_{\bar x}(-\infty), \label{Qop1}\\
\tilde \psi_{\bar x}(+\infty) = \mathcal Q_{\bar xx}\tilde{\psi}_x(-\infty) + \mathcal Q_{\bar x\bar x}\tilde{ \psi}_{\bar x}(-\infty). \label{Qop2}
\end{eqnarray}
It is possible to find the matrix elements of the operator $\mathcal
Q$ analytically in the limit of the weak pump pulse retaining only
the terms containing the combinations of $V_+$ and $V_+^*$ as:
\begin{eqnarray}
\label{Qvals}
\mathcal Q_{xx} &=& 1 -\frac{1}{2\hbar^2} \int_{-\infty}^{\infty} \mathrm dt \int_{-\infty}^t\mathrm dt' V_+^*(t)V_+(t') \mathrm e^{\mathrm i (\frac{\Omega_{\rm L}}{2}-\omega_0)(t-t')},\\
\mathcal Q_{\bar x\bar x} &=& 1 -\frac{1}{2\hbar^2} \int_{-\infty}^{\infty} \mathrm dt \int_{-\infty}^t\mathrm dt' V_+^*(t)V_+(t') \mathrm e^{\mathrm i (\frac{\Omega_{\rm L}}{2}+\omega_0)(t'-t)},\nonumber\\
\mathcal Q_{x\bar x}&=& -\frac{1}{2\hbar^2} \int_{-\infty}^{\infty} \mathrm dt \int_{-\infty}^t\mathrm dt' V_+^*(t)V_+(t') \mathrm e^{-\mathrm i \omega_0(t-t') + \mathrm i \frac{\Omega_{\rm L}(t+t')}{2}},\nonumber\\
\mathcal Q_{\bar x x}&=& -\frac{1}{2\hbar^2} \int_{-\infty}^{\infty} \mathrm dt \int_{-\infty}^t \mathrm dt' V_+^*(t)V_+(t') \mathrm e^{-\mathrm i \omega_0(t-t') - \mathrm i \frac{\Omega_{\rm L}(t+t')}{2}}.\nonumber
\end{eqnarray}
Equations~\eqref{Qop1}, \eqref{Qop2} together with the
expressions~\eqref{Qvals} determine the change of electron spin
caused by the pump pulse application.
\end{widetext}

\section{Spectrally wide pulses} \label{app:wide}

In the main text we presented theoretical results for
``Fourier-limited'' pulse, whose envelope function was exponential
and its spectral width $\Delta\omega$ is related to the pulse
duration $\tau_{\rm p}$ ($\rm p=\rm pump$ or $\rm probe$) by $\Delta \omega
= \tau_{\rm p}^{-1}$. While in the experiment the 2~ps
pulses are Fourier-limited, this is not the case for the longer
pulses. Still, the spectral linewidth becomes the narrower, the
longer the pulse duration is. To analyze whether this might have an
impact on the experimental findings, we will briefly discuss the
situation for non-Fourier limited pulses. Our main result is
  as follows:
We find that there are changes on a quantitative level, but
qualitatively the scenario on the magnetic field dependence for
varying pulse durations remains the same.

\begin{figure}[hptb]
\includegraphics[width=0.5\textwidth]{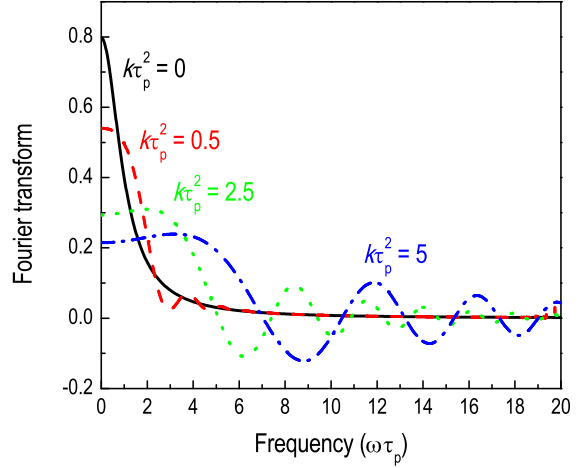}
\caption{Fourier transform of the pulses given by
Eq.~\eqref{chriped} for different values of the parameter
$k\tau_{\rm p}^2=0$ (black/solid), 0.5 (red/dashed), 2.5 (blue/dotted)
and 5 (green/dash-dotted)
curves, respectively. }\label{fig:chirped:tr}
\end{figure}

As an example we consider chirped pulses with an envelope function
\begin{equation}
\label{chriped}
f_{\rm p} = \cos{\left(k t^2\right)} \mathrm e^{-|t|/\tau_{\rm p}}.
\end{equation}
The duration of the pulse is still determined by $\tau_{\rm p}$, but
its spectral width is controlled by the independent parameter $k$.
The Fourier transforms of such pulses for different values of $k$
are presented in Fig.~\ref{fig:chirped:tr}. An increase of $k$ leads
to an increase of the spectral width of the pulse  roughly
proportional to $k\tau_{\rm p}$ while its amplitude drops because the same
power is distributed over a wider frequency range.

\begin{figure}[hptb]
\includegraphics[width=0.5\textwidth]{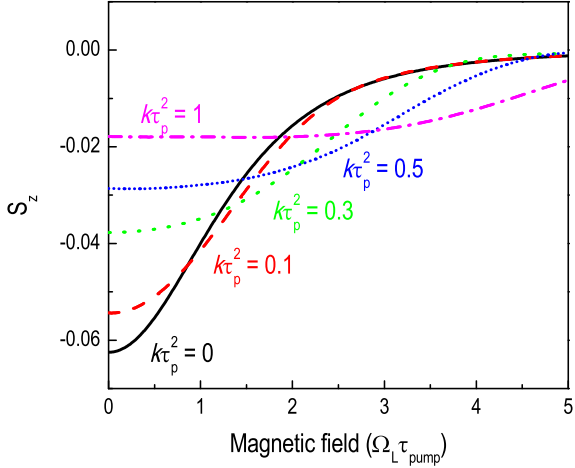}
\caption{Electron spin $z$ component generated by a single pump
pulse as function of magnetic field $\Omega_{\rm L} \tau_{\rm
pump}$. Different curves correspond to different spectral widths of
the pulse determined by the parameter $k\tau_{\rm pump}^2=0$
(black/solid), 0.1 (red/dashed), 0.3 (green/dotted), 0.5 (blue/short dotted),
and 1 (magenta/dash dotted).}\label{fig:chirped:S}
\end{figure}

Spectral broadening of the pulses results in modifications of the
spin coherence initialization and detection.
Figure~\ref{fig:chirped:S} shows the electron spin component after a
single pump pulse, $S_z^+$, as function of magnetic field for
different values of the parameter $k$. Spectral broadening of the
pulse results in a decrease of the electron spin $z$ component for
$\Omega\tau_{\rm pump}=0$ and in a weaker dependence on the magnetic
field. As a result, at relatively high magnetic fields the spin
coherence generation is more efficient for spectrally wide pulses as
compared with Fourier limited pulses because the coherent
superposition of the electron spin states is excited more
efficiently.

\begin{figure}[hptb]
\includegraphics[width=0.45\textwidth]{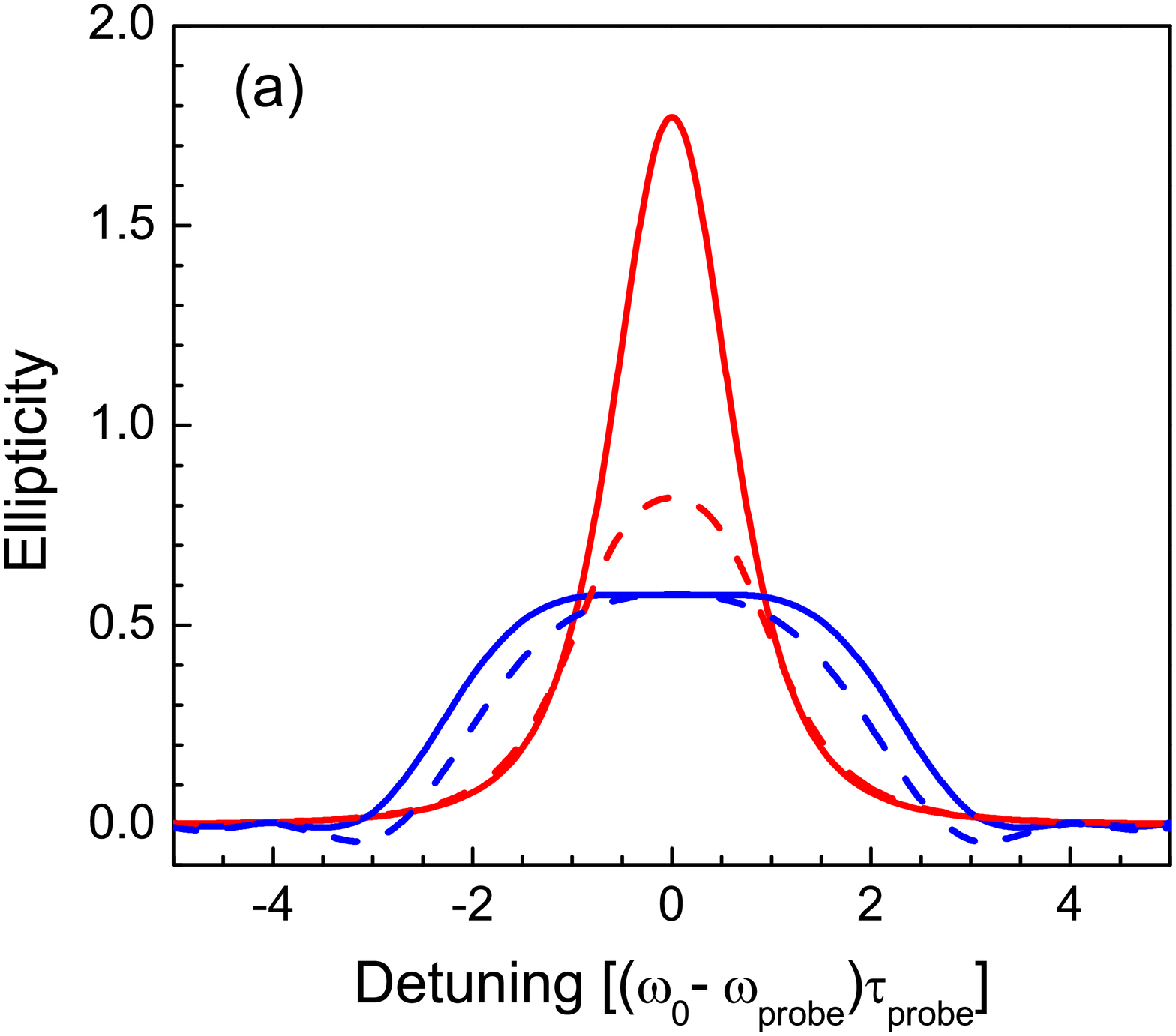}\\
\includegraphics[width=0.45\textwidth]{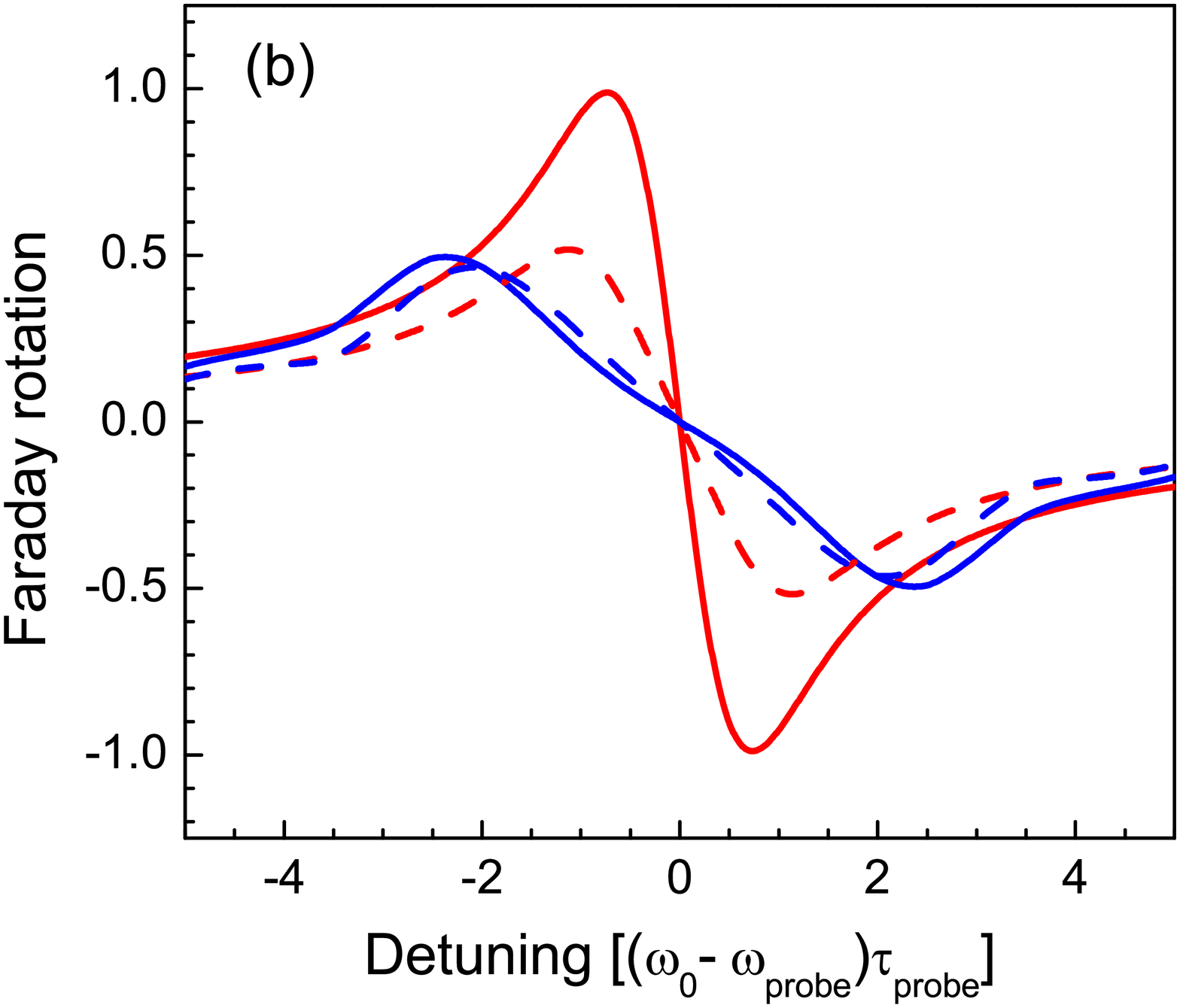}
\caption{Ellipticity (top) and Faraday rotation (bottom) signals as
function of detuning between the QD resonance and the probe optical
frequency. The red curves are calculated for $k\tau_{\rm p}^2=0$ and the
blue ones for $k\tau_{\rm p}^2=1$. The solid and dashed sets of curves
correspond to two values of magnetic field $\Omega_{\rm L}\tau_{\rm
probe}=1.5$ and $0.5$, respectively. The spin $z$ component is the
same for all curves. The signals are given in arbitrary
units.}\label{fig:signals:chirped}
\end{figure}

The spectral dependencies of the ellipticity and Faraday rotation
signals are shown in Fig.~\ref{fig:signals:chirped}. The red curves
correspond to $k\tau_{\rm p}^2=0$ and the blue ones to $k\tau_{\rm p}^2=1$. For
spectrally wide pulse the sensitivities of ellipticity and Faraday
rotation do not depend much on the magnetic field, since the solid
and dashed curves, corresponding to different values of the field
strength, almost coincide. By contrast, the magnetic field effect is
strong for spectrally narrow pulses. The appearance of the
oscillations in the spectral dependencies for broad pulses are
related with the specifics of the pulse Fourier transforms, see
oscillations in Fig.~\ref{fig:chirped:tr}.

Another reason for the spectral broadening of the pump/probe pulses
may be related with the jitter of the laser optical frequency. In
this case, each pulse generated by the laser is Fourier limited, but
its central frequency changes randomly from pulse to pulse. In this
case, the theory developed in Sec.~\ref{sec:gen} remains valid, but
the results should be averaged over the distribution of the optical
frequencies of the laser. Therefore we can safely conclude that the
results as derived in the main section will not change for
spectrally broadened pulses.

\end{document}